\documentclass[10pt,conference]{IEEEtran}
\IEEEoverridecommandlockouts
\usepackage{cite}
\usepackage{amsmath,amssymb,amsfonts}
\usepackage{algorithmic}
\usepackage{graphicx}
\usepackage{textcomp}
\usepackage{xcolor}
\usepackage{booktabs}
\usepackage{tabularx}
\usepackage{array}  
\usepackage{tcolorbox}
\usepackage{enumitem}

\usepackage{soulutf8}

\def\BibTeX{{\rm B\kern-.05em{\sc i\kern-.025em b}\kern-.08em
    T\kern-.1667em\lower.7ex\hbox{E}\kern-.125emX}}
\begin{document}

\title{Improving Automated Secure Code Reviews: \\A Synthetic Dataset for Code Vulnerability Flaws
}

\author{
\IEEEauthorblockN{
Leonardo Centellas-Claros, Juan J. Alonso-Lecaros,\\
Juan Pablo Sandoval Alcocer, Andres Neyem
}
\IEEEauthorblockA{
\textit{Department of Computer Science, School of Engineering, Faculty of Engineering,}\\
\textit{Pontificia Universidad Católica de Chile, Santiago, Chile} \\
\{lcentellas, jalonsol, juanpablo.sandoval, aneyem\}@uc.cl
}
}

\maketitle

\begin{abstract}
Automation of code reviews using AI models has garnered substantial attention in the software engineering community as a strategy to reduce the cost and effort associated with traditional peer review processes. These models are typically trained on extensive datasets of real-world code reviews that address diverse software development concerns, including testing, refactoring, bug fixes, performance optimization, and maintainability improvements. 
However, a notable limitation of these datasets is the under representation of code vulnerabilities, critical flaws that pose significant security risks, with security-focused reviews comprising a small fraction of the data. 
This scarcity of vulnerability-specific data restricts the effectiveness of AI models in identifying and commenting on security-critical code. 
To address this issue, we propose the creation of a synthetic dataset consisting of vulnerability-focused reviews that specifically comment on security flaws. Our approach leverages Large Language Models (LLMs) to generate human-like code review comments for vulnerabilities, using insights derived from code differences and commit messages.
To evaluate the usefulness of the generated synthetic dataset, we plan to use it to fine-tune three existing code review models. We anticipate that the synthetic dataset will improve the performance of the original code review models.
\end{abstract}


\begin{IEEEkeywords}
Code Review Automation, Review Comments, Large Language Models, Synthetic Dataset Generation
\end{IEEEkeywords}

\section{Introduction}

Code review is a standard practice in modern software development, widely adopted in both commercial and open source projects. Despite its benefits, the process is time-consuming and resource-intensive. A study of eight popular software projects revealed that the median waiting time for the first response to a pull request ranges from 42 to 126 hours \cite{MiningCRDataWaitingTimes}. In addition, developers report dedicating a median of six hours per week to code review activities \cite{ImpactOfCodeReview}. These prolonged review times have led to the development of techniques to automate code review tasks \cite{AUGER, LLaMA-Reviewer, D-ACT}, with the aim of improving efficiency and reducing the workload of developers.

Among the various automation efforts, two tasks have received significant research attention: \textit{code-to-comment} and \textit{code \& comment-to-code}. The first involves generating review comments for submitted code and simulating the reasoning process of human reviewers. The second focuses on implementing automatic code changes based on both the submitted code and the corresponding review comments. These approaches rely on large, high-quality datasets for training models, which often require thousands of labeled instances \cite{StateOfTheArt}. Common data sources include platforms such as Gerrit \cite{D-ACT}, designed specifically for code reviews, and GitHub \cite{CodeEditor}.

The quality of the dataset is an important determinant of the effectiveness and precision of automated code review systems. A recent evaluation of a state-of-the-art tool found that up to 25\% of the data used for training contained noise \cite{StateOfTheArt}. Furthermore, while these tools provide general-purpose code review capabilities, they perform poorly on specific review tasks, such as simplifying return statements or modifying method visibility \cite{StateOfTheArt}. These findings underscore the need for higher-quality datasets to advance the field.

The challenges are particularly pronounced in the domain of security-related code reviews, where existing datasets often lack adequate representation of vulnerability-related reviews. For example, a recent study that analyzed a dataset of 20,000 code review comments identified only 614 as security related \cite{SecurityDefectDetection}. This imbalance highlights an important gap in the diversity of security issues captured. One major factor contributing to this limitation is the lack of sufficient security knowledge among developers, which has been identified as the primary challenge to ensuring effective security practices during code reviews \cite{SoftwareSecurityDuringModernCodeReview}. This deficiency not only skews the representation of security-related reviews, but also limits the ability of automated tools to generalize effectively, reducing their reliability in identifying and addressing critical vulnerabilities.

To address the need for larger and higher quality training data, we propose a novel approach to obtaining vulnerability code review data using the potential of Large Language Models (LLMs). Specifically, our goal is to use vulnerability-related commits, changes often prompted by various factors, not just code reviews, to generate synthetic code reviews. These synthetic reviews emulate the feedback that, in a real scenario, would have prompted the creation of these commits. We aim to generate a large dataset of synthetic vulnerability-related code reviews, overcoming the scarcity of such data in existing datasets. We believe that the LLM-based approach can produce human-like code reviews and generate a synthetic dataset capable of improving the performance of the current code-to-comment review models.

\section{Related Work}
This sections summarizes the related work regarding automatic code review, code review dataset, and artificial dataset generation.

\subsection{Automatic Code Review}
Numerous approaches have been proposed to automate various aspects of the code review process, many leveraging language models \cite{CodeReviewer, Pre-TrainedModelsToCodeReview, AUGER, LLaMA-Reviewer, D-ACT, ImprovingAutomatedCodeReviews}. 

Li et al. \cite{CodeReviewer} introduced CodeReviewer, a model designed to address three key tasks: assessing whether a code change is ready for acceptance, generating review comments for a code change (\textit{code-to-comment}), and proposing modifications to a code change based on a review (\textit{code \& comment-to-code}). Their approach emphasized extensive pre-training on code-related tasks to enable the model to capture relationships between code and comments effectively.

Similarly, Tufano et al. \cite{Pre-TrainedModelsToCodeReview} focused on the importance of pre-training tasks, ensuring that the resulting model possessed general knowledge of the programming languages relevant to the downstream tasks. This technique was applied to both the \textit{code-to-comment} and \textit{code \& comment-to-code} tasks. Their dataset included over 1.4 million samples for pre-training and approximately 160,000 for fine-tuning.

Li et al. \cite{AUGER} proposed another approach, AUGER, which generated review comments using data collected from GitHub. They enhanced their dataset through a simple data augmentation technique applied to review comments, creating additional samples with similar meanings to increase the dataset's diversity.

All the aforementioned techniques rely heavily on the size and quality of the datasets used. This reliance has driven the development of numerous approaches for collecting and curating code review data.

\subsection{Code Review Datasets}
Previous research has proposed several datasets to study code review tasks. Tufano et al. \cite{TowardsAutomatingCodeReview} introduced a dataset derived from Java projects, comprising approximately 17K samples. Each sample is represented as a tuple in the form of $\langle C_s, R_{\text{nl}} \rangle \to C_r$, where $C_s$ denotes the submitted code, $R_{\text{nl}}$ is the natural language review comment, and $C_r$ represents the revised code.

Li et al. \cite{CodeReviewer} constructed one of the largest datasets to date by mining data from 10K GitHub projects. For the task of \textit{code \& comment-to-code}, this dataset contains 150,000 training samples, similarly structured as $\langle C_s, R_{\text{nl}} \rangle \to C_r$, but with code represented as diff hunks, emphasizing changes rather than entire code snippets. Guo et al. \cite{ExploringThePotentialOfChatGPT} later published an updated and enlarged version of this dataset, aiming to improve its quality.

Liang et al. \cite{CuratedEmail-BasedCR} introduced a large-scale dataset by mining code review discussions conducted via email. This approach enabled them to uncover previously overlooked data, resulting in a new dataset derived from 167 open-source software (OSS) projects.

These studies highlight the significant efforts invested in gathering high-quality data and underscore the potential benefits of exploring new methods to obtain code review information.

While these studies provide large collections of general-purpose code review data, the proportion of security-related samples remains minimal. Reports indicate that security-focused reviews account for less than 4\% of the data in some datasets \cite{SecurityDefectDetection}.

\subsection{Artificial Dataset Generation}
With the growing demand for better and more powerful models, there has been an increasing need for larger datasets to train them. This has prompted the use of LLMs to generate datasets tailored to specific tasks. For instance, Abdullin et al. \cite{SyntheticDialogueDatasetGeneration} created a dataset by using two LLM agents to simulate dialogues between a human and a conversational bot. Similarly, Acharya et al. \cite{LLMBasedGenerationofItemDescription} employed an LLM to generate movie descriptions, providing an alternative to web scraping for recommender systems.

Honovich et al. \cite{UnnaturalInstructions} developed a large dataset of natural language instructions, demonstrating the effectiveness of models trained on synthetic data. Ye et al. \cite{ZeroGenEfficientZeroShotLearning} introduced a method to train models based on the outputs of another LLM, emphasizing the importance of task-specific prompts. Finally, Yu et al. \cite{LLMasAttributedTrainingDataGenerator} underscored the critical role of prompt design in determining the quality of generated data.

All these methods underscore the benefits of using LLMs to augment and create datasets with artificial samples. However, to the best of our knowledge, no study has investigated the ability of LLMs to reverse-engineer commits into review comments, especially for security-related commits. This study aims to bridge that gap and further demonstrate the potential of LLMs for artificial dataset augmentation.

\section{Methodology}
This study explores the feasibility of using large language models (LLMs) to emulate the code review process by generating a dataset of synthetic vulnerability-related code reviews. These reviews will be based on security-focused commit diffs and the corresponding commit messages. The overarching hypothesis is that \emph{LLMs can produce human-like code reviews from commits with high accuracy, creating a valuable resource for enhancing automated code-to-comment review models}.

\subsection{Research Questions}
We propose the following research questions.

\begin{itemize}
    \item \textbf{RQ1:} \emph{How accurate are LLMs in generating synthetic datasets of vulnerability code reviews from commits?}
    \item \textbf{RQ2:} \emph{Does the artificially generated dataset improve the precision of code review models?}
\end{itemize}

\textbf{RQ1} is focused on evaluating the ability of large language models (LLMs) to generate accurate synthetic datasets of vulnerability code reviews based on commits. To address this, we plan to assess the performance of various LLM models and explore the effect of different prompt engineering techniques on their output. This research question is important to understanding whether LLMs can produce high-quality, human-like reviews that effectively emulate real-world scenarios. By addressing this, our aim is to assess the feasibility of using LLMs to fill gaps in existing datasets, particularly in the context of security-focused reviews, where data scarcity limits the progress of automated code review models.

\textbf{RQ2} examines the impact of the artificially generated dataset on the precision of automated code review models. To address this, we plan to fine-tune code review models using the synthetic dataset generated by LLMs. The goal is to determine whether training models with our synthetic data can improve the  ability of this models in generating accurate and reliable reviews. This question is relevant because it connects the quality of the synthetic dataset with its practical applications, exploring whether our dataset can serve as a valuable resource to advance in automated code review generation, especially in scenarios involving security vulnerabilities.



\subsection{Synthetic Dataset Generation (RQ1)}
We propose a six-step methodology to generate and validate a synthetic dataset.

\subsubsection{Project selection}
Our first step involves selecting a representative set of open-source projects from GitHub. Given Java's prevalence in existing code review models, our research will focus exclusively on Java-based projects. To ensure the selection of mature and active repositories, we will adopt the criteria proposed by Tufano et al. \cite{TowardsAutomatingCodeReview}, considering only projects with at least 50 pull requests (PRs). This threshold provides a reliable foundation for our analysis by targeting repositories with significant development activity.

\subsubsection{Commit selection}
As our goal is to generate code reviews based on commit diffs and comments, we focus on commits with specific characteristics to ensure their suitability for emulating realistic reviews. Commits that involve changes to a single \texttt{.java} file are ideal candidates, as they typically represent small, self-contained, and focused modifications. This is particularly important for security code reviews, which often target specific sections of code where potential vulnerabilities may exist. By narrowing the scope to single-file changes, we increase the likelihood of producing accurate review comments based on source code changes.
To maintain relevance, we exclude certain types of commits. Merge commits, which primarily resolve conflicts, are excluded because they may introduce multiple types of changes, making them less suitable for focused analysis. These are identified by omitting commits with two parent commits, as flagged by GitHub. Additionally, we exclude commits that do not modify source code files, as such changes are not useful for generating actionable code reviews.

\subsubsection{Filtering vulnerability based code commits}
To identify whether a commit addresses a vulnerability, we will use a security keyword search method inspired by previous studies \cite{WhySecurityDefectsGoUnnoticed, CharacteristicsOfTheVulnerableCodeChanges}. We plan to start with a list of security-related keywords reported in prior research \cite{EmpiricalAnalysisOfSecurityRelated}. This list will be refined using the methodology proposed by Alfadel et al. \cite{EmpiricalAnalysisOfSecurityRelated}.

First, for each keyword, we will randomly select a statistically significant sample of commits that contain the said keyword, ensuring a 95\% confidence level and a 5\% confidence interval. Subsequently, we will choose a random sample of commits that do not contain any of the selected keywords. The size of this sample will also be calculated to maintain a confidence level of 95\%. Including this sample will help identify potential new keywords to expand our list.

The process will consist of two iterations. In the first iteration, two authors will independently review the sampled commits to determine whether they address vulnerabilities, resolving any disagreements with the help of a third author. Keywords with a precision of over 75\% will be retained, and new potential keywords identified during the review process will be noted. This threshold may be adjusted depending on the number of commits remaining after certain keywords are removed to maintain a sufficiently large and representative dataset.

In the second iteration, we will repeat the sampling process using the newly identified keywords. Two authors will again independently review the samples, retaining only keywords that demonstrate high precision. This iterative process will result in a refined list of security-related keywords, which we will then apply to all commits to filter those relevant to vulnerabilities.


\subsubsection{LLMs under analysis}
We plan to evaluate four publicly available large-language models (LLMs): GPT-4o from OpenAI, Claude 3.5 Sonnet from Anthropic - which reports superior performance compared to GPT-4o in code-related tasks - Flan-T5 from Google and Qwen 2.5. Notably, Claude 3.5 Sonnet and Qwen 2.5 are specifically designed to excel in code-related tasks, making them particularly relevant for our analysis.

\subsubsection{Prompt Design}
Using the commit message and diff information from each commit, the goal is to design a prompt that instructs the model to generate a review comment that could plausibly have prompted the creation of the commit, relying solely on this data.

The prompt provided to the LLM will include the \textit{Diff} and \textit{Commit Message} from the revised code after the commit addressed the vulnerability. It will also include instructions to generate a synthetic review in natural language that could lead to transitioning the code from its vulnerable state (prior to the commit) to its current post-commit state.

We plan to design prompts using three techniques informed by existing literature: zero-shot \cite{FewShot}, a straightforward prompting approach; Chain-of-Thought (CoT) \cite{CoT}, a technique designed to enhance the model's performance in tasks requiring complex reasoning; Self-Reflection \cite{Self-Reflection}, which enables the model to evaluate its initial response and refine it for better accuracy. These strategies were chosen as they align well with the task at hand, balancing simplicity, reasoning enhancement, and iterative improvement. To refine the prompts, we will select a random sample of 50 commits and use them to iteratively test and improve the prompts. This process will continue until all authors agree that the prompts consistently produce satisfactory responses. Below is an example of a zero-shot prompt we intend to use:

\vspace{0.2cm}

\texttt{Given this diff hunk "\{\{Diff\}\}" and this commit message "\{\{Message\}\}" belonging to a commit that addresses a vulnerability. Generate a code review that could have led to making said commit in the first place. Write it like a reviewer who found a vulnerability on the code.}
\vspace{0.2cm}

Here, \texttt{\{\{Diff\}\}} refers to the diff hunk generated by the security-related commit, and \texttt{\{\{Message\}\}} is the commit message authored by the commit creator.

\subsubsection{Prompt/LLM Evaluation}
To determine the most suitable LLM/prompt combination, we plan to select a random sample of 100 vulnerability commits obtained after Step 3. We will then use these as input for each Prompt/LLM combination. With four LLMs and three prompt strategies, this step will yield 1,200 code review comments.

To evaluate the precision of each combination, two authors will independently review each of the 1,200 generated code review comments and categorize each review as suitable or not. To determine suitability, each author will assess (1) whether the generated reviews are coherent, (2) whether they address the identified vulnerability, and (3) whether they could plausibly prompt the corresponding commit. We have decided to analyze only 100 commits to make manual review feasible, as it results in 1,200 code reviews.

We will measure the agreement between the authors using Cohen's Kappa to evaluate the consistency of their evaluations. Any discrepancies will be resolved collaboratively. This process will enable us to measure the precision of each prompt strategy and LLM, answer RQ1, and identify which prompt and LLM combination is the most accurate for this task.

\subsubsection{Dataset Generation}
After determining which prompt/LLM combination yielded the best results in the previous step, we will generate the synthetic dataset using all the commits found in Step 3 as input for the prompt/LLM combination identified in Step 6.

\subsection{Evaluating the Synthetic Dataset (RQ2)}

To evaluate the usefulness of the generated dataset and address the second research question, we design a five-step methodology.

\subsubsection{Baseline for Comparison}
We plan to use the following models as baseline for comparison:
\begin{itemize}
    \item CodeReviewer
\cite{CodeReviewer}, a widely used baseline in various studies \cite{LLaMA-Reviewer, StateOfTheArt}, which includes a well-documented artifact and a publicly available pre-trained version.

    \item A more recent fine-tuned version of CodeReviewer, proposed by Lin et al., which has demonstrated the ability to generate more meaningful review comments \cite{ImprovingAutomatedCodeReviews}.

    \item GPT-4o as a representative of LLMs with the prompt used on \cite{LLaMA-Reviewer} for review comment generation.
\end{itemize}


\subsubsection{Model fine-tuning}
To assess the impact of the generated synthetic dataset, we plan to use it to fine-tune the following models and compare their performance against the baseline models. 
\begin{itemize}
    \item The pre-trained model of CodeReviewer.
    \item The final version of CodeReviewer trained on its \textit{code-to-comment} dataset.
    \item The model proposed by Lin et al. with the oversampled dataset from CodeReviewer.
\end{itemize}
This combinations will help evaluate the usefulness of the dataset and its influence over another dataset.
As the primary goal of this study is to evaluate the usability of a synthetic dataset, we do not intend to perform an in-depth analysis of hyperparameters. Instead, we will use the same hyperparameters as those employed in the original CodeReviewer fine-tuning tasks.

\subsubsection{Testing Dataset}
To accurately measure the performance of the resulting model in real-life scenarios, a dataset other than the synthetic one will be required. Given the scarcity of a dedicated security-specific code review dataset for Java, we plan to use a filtered subset of the test partition from the dataset proposed by Li et al. \cite{CodeReviewer}. Specifically, we will reuse the keyword list obtained in the third step of our synthetic dataset generation process to filter security-related review data from their test partition.

Instead of filtering by commit messages, we will filter based on the presence of security-related keywords within the review comments. Each sample will then be manually reviewed by two authors to ensure accuracy and relevance. This approach will provide an evaluation dataset composed of real-world security-focused review data, allowing a fair comparison between models on unseen data on which none of them have been trained.

Although this process is expected to yield a limited number of samples, we anticipate creating a small but valuable dataset of security-focused code reviews. This dataset will facilitate an equitable evaluation of the models using actual data. In our initial filtering of the test datasets for CodeReviewer, using the preliminary list of security-related keywords, we identified 43 and 63 potentially security-related samples in the test partitions for the \textit{code-to-comment} and \textit{code \& comment-to-code} tasks, respectively. In particular, only three samples overlapped between the two datasets.

\subsubsection{Performance metrics}
Similar to the approach by Li et al. \cite{CodeReviewer}, we intend to use BLEU-4 as a primary evaluation metric. Furthermore, we plan to manually evaluate all the reviews generated based on the metrics proposed by Lin et al. \cite{ImprovingAutomatedCodeReviews}, adapted to our context: (a) \emph{Semantic Equivalence}: The generated review conveys the same meaning as the ground truth; and (b) \emph{Applicability}: The generated comment raises a valid suggestion or concern specifically addressing the identified vulnerability. All manual evaluations will be independently conducted by two authors, with a kappa score reported to measure inter-rater agreement.

\subsubsection{Evaluation}
Using the security dataset from real samples obtained in step three, we plan to generate review comments with the fine-tuned models and the baselines. We will then compare the generated comments against the ground truth using the defined evaluation metrics. This process will address RQ2, demonstrating the usability of a synthetic dataset.

\section{Early Findings}
Following the first step described in the methodology, we identified 5,973 candidate repositories. From these, we mined a total of 3,827,517 potentially usable commits.

Using the keyword list from Alfadel et al. \cite{EmpiricalAnalysisOfSecurityRelated}, we filtered the commits and identified 43,131 potentially relevant ones. After further discussion, we decided to exclude any commit where the modified filename contained the substring "test," as flaws in test files do not pose a direct security threat and could introduce unnecessary noise into the dataset. This filtering process reduced the total to 35,950 potentially security-related commits.

If most or all keywords from the initial list are retained in the refined list, we could potentially create a substantial vulnerability-focused code review dataset.

\section{Threats to Validity}
This section describes the possible threats to validity that could impact the study.

\textit{Internal Validity}
The evaluation of the synthetic reviews is inherently subjective, as it relies on the authors' judgment to determine the accuracy of the generated review comments. To reduce bias, the evaluation process involves two authors independently assessing the reviews, with disagreements resolved collaboratively.

The security filtering process is based on a predefined list of keywords from prior literature. Although we plan to refine this list, it may still exclude some security-relevant commits that are not captured by the selected keywords.

The iterative prompt refinement process, using a sample of 50 commits, may introduce bias, as the sample might not fully represent the diversity of commits in the dataset. To address this, we plan to use a different sample for each iteration to increase coverage and reduce potential bias.

\textit{External Validity}
The study’s focus on Java projects inherently limits the generalizability of its findings to other programming languages. Language-specific practices and features may influence the synthetic dataset generation process and its broader applicability. To address this, Java was chosen due to its prominence as one of the most widely used languages in code review research, ensuring relevance to a substantial portion of the domain.

Additionally, the study targets vulnerability-fixing commits—a specific subset of code review scenarios—due to the critical importance of security-focused reviews and their underrepresentation in existing datasets. This targeted focus ensures the methodology tackles a well-defined problem, while also providing a foundation for future research to extend the approach to other review types.

\section{Summary}
This paper describes a plan to develop a novel technique for gathering code review data, with a focus on creating a new dataset for vulnerability code reviews—an essential but underrepresented type of review. By leveraging the capabilities of general-purpose LLMs, we aim to generate human-like reviews from security-related commits. This approach seeks to address the demand for high-quality datasets by utilizing previously overlooked commits, paving the way for advancements in automated code review research.

\section*{Acknowledgments}

Leonardo Centellas-Claros is supported by the Chilean National Agency for Research and Development (ANID)/Scholarship Program/DOCTORADO NACIONAL/ 2024-21240734. Andres Neyem and Leonardo Centellas-Claros acknowledge support from the National Center for Artificial Intelligence (CENIA), Grant/Award Number: BASAL, ANID, FB210017. Juan Pablo Sandoval Alcocer thanks the Programa de Inserción Académica 2022, Vicerrectoría Académica y Prorrectoría, at the Pontificia Universidad Católica de Chile.

\clearpage

\bibliographystyle{IEEEtran}
\bibliography{references}

\begin{thebibliography}{10}
\providecommand{\url}[1]{#1}
\csname url@samestyle\endcsname
\providecommand{\newblock}{\relax}
\providecommand{\bibinfo}[2]{#2}
\providecommand{\BIBentrySTDinterwordspacing}{\spaceskip=0pt\relax}
\providecommand{\BIBentryALTinterwordstretchfactor}{4}
\providecommand{\BIBentryALTinterwordspacing}{\spaceskip=\fontdimen2\font plus
\BIBentryALTinterwordstretchfactor\fontdimen3\font minus \fontdimen4\font\relax}
\providecommand{\BIBforeignlanguage}[2]{{%
\expandafter\ifx\csname l@#1\endcsname\relax
\typeout{** WARNING: IEEEtran.bst: No hyphenation pattern has been}%
\typeout{** loaded for the language `#1'. Using the pattern for}%
\typeout{** the default language instead.}%
\else
\language=\csname l@#1\endcsname
\fi
#2}}
\providecommand{\BIBdecl}{\relax}
\BIBdecl

\bibitem{MiningCRDataWaitingTimes}
\BIBentryALTinterwordspacing
G.~Kudrjavets, A.~Kumar, N.~Nagappan, and A.~Rastogi, ``Mining code review data to understand waiting times between acceptance and merging: an empirical analysis,'' in \emph{Proceedings of the 19th International Conference on Mining Software Repositories}, ser. MSR '22.\hskip 1em plus 0.5em minus 0.4em\relax New York, NY, USA: Association for Computing Machinery, 2022, p. 579–590. [Online]. Available: \url{https://doi.org/10.1145/3524842.3528432}
\BIBentrySTDinterwordspacing

\bibitem{ImpactOfCodeReview}
A.~Bosu and J.~C. Carver, ``Impact of peer code review on peer impression formation: A survey,'' in \emph{2013 ACM / IEEE International Symposium on Empirical Software Engineering and Measurement}, 2013, pp. 133--142.

\bibitem{AUGER}
\BIBentryALTinterwordspacing
L.~Li, L.~Yang, H.~Jiang, J.~Yan, T.~Luo, Z.~Hua, G.~Liang, and C.~Zuo, ``Auger: automatically generating review comments with pre-training models,'' in \emph{Proceedings of the 30th ACM Joint European Software Engineering Conference and Symposium on the Foundations of Software Engineering}, ser. ESEC/FSE 2022.\hskip 1em plus 0.5em minus 0.4em\relax New York, NY, USA: Association for Computing Machinery, 2022, p. 1009–1021. [Online]. Available: \url{https://doi.org/10.1145/3540250.3549099}
\BIBentrySTDinterwordspacing

\bibitem{LLaMA-Reviewer}
\BIBentryALTinterwordspacing
J.~Lu, L.~Yu, X.~Li, L.~Yang, and C.~Zuo, ``{ LLaMA-Reviewer: Advancing Code Review Automation with Large Language Models through Parameter-Efficient Fine-Tuning },'' in \emph{2023 IEEE 34th International Symposium on Software Reliability Engineering (ISSRE)}.\hskip 1em plus 0.5em minus 0.4em\relax Los Alamitos, CA, USA: IEEE Computer Society, Oct. 2023, pp. 647--658. [Online]. Available: \url{https://doi.ieeecomputersociety.org/10.1109/ISSRE59848.2023.00026}
\BIBentrySTDinterwordspacing

\bibitem{D-ACT}
C.~Pornprasit, C.~Tantithamthavorn, P.~Thongtanunam, and C.~Chen, ``D-act: Towards diff-aware code transformation for code review under a time-wise evaluation,'' in \emph{2023 IEEE International Conference on Software Analysis, Evolution and Reengineering (SANER)}, 2023, pp. 296--307.

\bibitem{StateOfTheArt}
\BIBentryALTinterwordspacing
R.~Tufano, O.~Dabi\'{c}, A.~Mastropaolo, M.~Ciniselli, and G.~Bavota, ``Code review automation: Strengths and weaknesses of the state of the art,'' \emph{IEEE Trans. Softw. Eng.}, vol.~50, no.~2, p. 338–353, Jan. 2024. [Online]. Available: \url{https://doi.org/10.1109/TSE.2023.3348172}
\BIBentrySTDinterwordspacing

\bibitem{CodeEditor}
\BIBentryALTinterwordspacing
J.~Li, G.~Li, Z.~Li, Z.~Jin, X.~Hu, K.~Zhang, and Z.~Fu, ``Codeeditor: Learning to edit source code with pre-trained models,'' \emph{ACM Trans. Softw. Eng. Methodol.}, vol.~32, no.~6, Sep. 2023. [Online]. Available: \url{https://doi.org/10.1145/3597207}
\BIBentrySTDinterwordspacing

\bibitem{SecurityDefectDetection}
\BIBentryALTinterwordspacing
J.~Yu, L.~Fu, P.~Liang, A.~Tahir, and M.~Shahin, ``{ Security Defect Detection via Code Review: A Study of the OpenStack and Qt Communities },'' in \emph{2023 ACM/IEEE International Symposium on Empirical Software Engineering and Measurement (ESEM)}.\hskip 1em plus 0.5em minus 0.4em\relax Los Alamitos, CA, USA: IEEE Computer Society, Oct. 2023, pp. 1--12. [Online]. Available: \url{https://doi.ieeecomputersociety.org/10.1109/ESEM56168.2023.10304852}
\BIBentrySTDinterwordspacing

\bibitem{SoftwareSecurityDuringModernCodeReview}
\BIBentryALTinterwordspacing
L.~Braz and A.~Bacchelli, ``Software security during modern code review: the developer’s perspective,'' in \emph{Proceedings of the 30th ACM Joint European Software Engineering Conference and Symposium on the Foundations of Software Engineering}, ser. ESEC/FSE 2022.\hskip 1em plus 0.5em minus 0.4em\relax New York, NY, USA: Association for Computing Machinery, 2022, p. 810–821. [Online]. Available: \url{https://doi.org/10.1145/3540250.3549135}
\BIBentrySTDinterwordspacing

\bibitem{CodeReviewer}
\BIBentryALTinterwordspacing
Z.~Li, S.~Lu, D.~Guo, N.~Duan, S.~Jannu, G.~Jenks, D.~Majumder, J.~Green, A.~Svyatkovskiy, S.~Fu, and N.~Sundaresan, ``Automating code review activities by large-scale pre-training,'' in \emph{Proceedings of the 30th ACM Joint European Software Engineering Conference and Symposium on the Foundations of Software Engineering}, ser. ESEC/FSE 2022.\hskip 1em plus 0.5em minus 0.4em\relax New York, NY, USA: Association for Computing Machinery, 2022, p. 1035–1047. [Online]. Available: \url{https://doi.org/10.1145/3540250.3549081}
\BIBentrySTDinterwordspacing

\bibitem{Pre-TrainedModelsToCodeReview}
\BIBentryALTinterwordspacing
R.~Tufano, S.~Masiero, A.~Mastropaolo, L.~Pascarella, D.~Poshyvanyk, and G.~Bavota, ``Using pre-trained models to boost code review automation,'' in \emph{Proceedings of the 44th International Conference on Software Engineering}, ser. ICSE '22.\hskip 1em plus 0.5em minus 0.4em\relax New York, NY, USA: Association for Computing Machinery, 2022, p. 2291–2302. [Online]. Available: \url{https://doi.org/10.1145/3510003.3510621}
\BIBentrySTDinterwordspacing

\bibitem{ImprovingAutomatedCodeReviews}
\BIBentryALTinterwordspacing
H.~Y. Lin, P.~Thongtanunam, C.~Treude, and W.~Charoenwet, ``Improving automated code reviews: Learning from experience,'' in \emph{Proceedings of the 21st International Conference on Mining Software Repositories}, ser. MSR '24.\hskip 1em plus 0.5em minus 0.4em\relax New York, NY, USA: Association for Computing Machinery, 2024, p. 278–283. [Online]. Available: \url{https://doi.org/10.1145/3643991.3644910}
\BIBentrySTDinterwordspacing

\bibitem{TowardsAutomatingCodeReview}
\BIBentryALTinterwordspacing
R.~Tufano, L.~Pascarella, M.~Tufano, D.~Poshyvanyk, and G.~Bavota, ``Towards automating code review activities,'' in \emph{Proceedings of the 43rd International Conference on Software Engineering}, ser. ICSE '21.\hskip 1em plus 0.5em minus 0.4em\relax IEEE Press, 2021, p. 163–174. [Online]. Available: \url{https://doi.org/10.1109/ICSE43902.2021.00027}
\BIBentrySTDinterwordspacing

\bibitem{ExploringThePotentialOfChatGPT}
\BIBentryALTinterwordspacing
Q.~Guo, J.~Cao, X.~Xie, S.~Liu, X.~Li, B.~Chen, and X.~Peng, ``Exploring the potential of chatgpt in automated code refinement: An empirical study,'' in \emph{Proceedings of the IEEE/ACM 46th International Conference on Software Engineering}, ser. ICSE '24.\hskip 1em plus 0.5em minus 0.4em\relax New York, NY, USA: Association for Computing Machinery, 2024. [Online]. Available: \url{https://doi.org/10.1145/3597503.3623306}
\BIBentrySTDinterwordspacing

\bibitem{CuratedEmail-BasedCR}
\BIBentryALTinterwordspacing
M.~Liang, W.~Charoenwet, and P.~Thongtanunam, ``Curated email-based code reviews datasets,'' in \emph{Proceedings of the 21st International Conference on Mining Software Repositories}, ser. MSR '24.\hskip 1em plus 0.5em minus 0.4em\relax New York, NY, USA: Association for Computing Machinery, 2024, p. 294–298. [Online]. Available: \url{https://doi.org/10.1145/3643991.3644872}
\BIBentrySTDinterwordspacing

\bibitem{SyntheticDialogueDatasetGeneration}
\BIBentryALTinterwordspacing
Y.~Abdullin, D.~Molla-Aliod, B.~Ofoghi, J.~Yearwood, and Q.~Li, ``Synthetic dialogue dataset generation using llm agents,'' 2024. [Online]. Available: \url{https://arxiv.org/abs/2401.17461}
\BIBentrySTDinterwordspacing

\bibitem{LLMBasedGenerationofItemDescription}
\BIBentryALTinterwordspacing
A.~Acharya, B.~Singh, and N.~Onoe, ``Llm based generation of item-description for recommendation system,'' in \emph{Proceedings of the 17th ACM Conference on Recommender Systems}, ser. RecSys '23.\hskip 1em plus 0.5em minus 0.4em\relax New York, NY, USA: Association for Computing Machinery, 2023, p. 1204–1207. [Online]. Available: \url{https://doi.org/10.1145/3604915.3610647}
\BIBentrySTDinterwordspacing

\bibitem{UnnaturalInstructions}
\BIBentryALTinterwordspacing
O.~Honovich, T.~Scialom, O.~Levy, and T.~Schick, ``Unnatural instructions: Tuning language models with (almost) no human labor,'' in \emph{Proceedings of the 61st Annual Meeting of the Association for Computational Linguistics (Volume 1: Long Papers)}, A.~Rogers, J.~Boyd-Graber, and N.~Okazaki, Eds.\hskip 1em plus 0.5em minus 0.4em\relax Toronto, Canada: Association for Computational Linguistics, Jul. 2023, pp. 14\,409--14\,428. [Online]. Available: \url{https://aclanthology.org/2023.acl-long.806}
\BIBentrySTDinterwordspacing

\bibitem{ZeroGenEfficientZeroShotLearning}
\BIBentryALTinterwordspacing
J.~Ye, J.~Gao, Q.~Li, H.~Xu, J.~Feng, Z.~Wu, T.~Yu, and L.~Kong, ``Zerogen: Efficient zero-shot learning via dataset generation,'' 2022. [Online]. Available: \url{https://arxiv.org/abs/2202.07922}
\BIBentrySTDinterwordspacing

\bibitem{LLMasAttributedTrainingDataGenerator}
Y.~Yu, Y.~Zhuang, J.~Zhang, Y.~Meng, A.~Ratner, R.~Krishna, J.~Shen, and C.~Zhang, ``Large language model as attributed training data generator: a tale of diversity and bias,'' in \emph{Proceedings of the 37th International Conference on Neural Information Processing Systems}, ser. NIPS '23.\hskip 1em plus 0.5em minus 0.4em\relax Red Hook, NY, USA: Curran Associates Inc., 2024.

\bibitem{WhySecurityDefectsGoUnnoticed}
R.~Paul, A.~K. Turzo, and A.~Bosu, ``Why security defects go unnoticed during code reviews? a case-control study of the chromium os project,'' in \emph{2021 IEEE/ACM 43rd International Conference on Software Engineering (ICSE)}, 2021, pp. 1373--1385.

\bibitem{CharacteristicsOfTheVulnerableCodeChanges}
\BIBentryALTinterwordspacing
A.~Bosu, ``Characteristics of the vulnerable code changes identified through peer code review,'' in \emph{Companion Proceedings of the 36th International Conference on Software Engineering}, ser. ICSE Companion 2014.\hskip 1em plus 0.5em minus 0.4em\relax New York, NY, USA: Association for Computing Machinery, 2014, p. 736–738. [Online]. Available: \url{https://doi.org/10.1145/2591062.2591200}
\BIBentrySTDinterwordspacing

\bibitem{EmpiricalAnalysisOfSecurityRelated}
\BIBentryALTinterwordspacing
M.~Alfadel, N.~A. Nagy, D.~E. Costa, R.~Abdalkareem, and E.~Shihab, ``Empirical analysis of security-related code reviews in npm packages,'' \emph{Journal of Systems and Software}, vol. 203, p. 111752, 2023. [Online]. Available: \url{https://www.sciencedirect.com/science/article/pii/S0164121223001474}
\BIBentrySTDinterwordspacing

\bibitem{FewShot}
T.~B. Brown, B.~Mann, N.~Ryder, M.~Subbiah, J.~Kaplan, P.~Dhariwal, A.~Neelakantan, P.~Shyam, G.~Sastry, A.~Askell, S.~Agarwal, A.~Herbert-Voss, G.~Krueger, T.~Henighan, R.~Child, A.~Ramesh, D.~M. Ziegler, J.~Wu, C.~Winter, C.~Hesse, M.~Chen, E.~Sigler, M.~Litwin, S.~Gray, B.~Chess, J.~Clark, C.~Berner, S.~McCandlish, A.~Radford, I.~Sutskever, and D.~Amodei, ``Language models are few-shot learners,'' in \emph{Proceedings of the 34th International Conference on Neural Information Processing Systems}, ser. NIPS '20.\hskip 1em plus 0.5em minus 0.4em\relax Red Hook, NY, USA: Curran Associates Inc., 2020.

\bibitem{CoT}
J.~Wei, X.~Wang, D.~Schuurmans, M.~Bosma, B.~Ichter, F.~Xia, E.~H. Chi, Q.~V. Le, and D.~Zhou, ``Chain-of-thought prompting elicits reasoning in large language models,'' in \emph{Proceedings of the 36th International Conference on Neural Information Processing Systems}, ser. NIPS '22.\hskip 1em plus 0.5em minus 0.4em\relax Curran Associates Inc., 2024.

\bibitem{Self-Reflection}
\BIBentryALTinterwordspacing
M.~Renze and E.~Guven, ``Self-reflection in llm agents: Effects on problem-solving performance,'' 2024. [Online]. Available: \url{https://arxiv.org/abs/2405.06682}
\BIBentrySTDinterwordspacing

\end{thebibliography}

\end{document}